\documentclass[aps,prd,twocolumn,showpacs,superscriptaddress,nofootinbib]{revtex4-1}

\usepackage[pdftex]{graphicx}

\usepackage{amsmath}
\usepackage[latin1]{inputenc}
\usepackage{txfonts}
\usepackage{hyperref}

\usepackage{xcolor}

\begin{document}

\title{Optimal synchronization of Kuramoto oscillators: a dimensional reduction approach }
\author{Rafael S. Pinto}
\email{rsoaresp@gmail.com}
\affiliation{Instituto de F\'\i sica ``Gleb Wataghin'', UNICAMP, 13083-859 Campinas, SP, Brazil.}
\author{Alberto Saa}
\email{asaa@ime.unicamp.br}
\affiliation{
Departamento de Matem\'atica Aplicada, 
 UNICAMP,  13083-859 Campinas, SP, Brazil.}

\date{\today}

\begin{abstract}
A recently proposed dimensional reduction approach  for studying synchronization in the Kuramoto model  
 is employed to build  optimal network topologies to favor or to suppress synchronization. The approach is based in the introduction
 of a collective coordinate for the time evolution of the phase locked oscillators, in the spirit of the Ott-Antonsen ansatz.
 We show that the optimal synchronization of a Kuramoto network demands the
maximization of the quadratic function $\omega^T L \omega$, where $\omega$ stands for the vector of 
the
 natural frequencies of the oscillators, and $L$ for the network Laplacian matrix. Many recently obtained numerical results can be re-obtained analytically
 and in a simpler way from our maximization  condition. A computationally  efficient {hill climb} rewiring algorithm is proposed to generate 
 networks with optimal synchronization properties. Our approach can be easily adapted  to the case of the Kuramoto models with both attractive and repulsive  interactions, and again many   recent numerical results can be rederived in a simpler and clearer analytical manner.
 \end{abstract}
\pacs{89.75.Fb, 05.45.Xt, 89.75.Hc}
\maketitle

\section{Introduction}

Synchronization  \cite{strogatz2003} is present in a myriad of natural and synthetic systems, ranging from metabolic processes in populations of yeast \cite{monte2007} to power grids \cite{motter2013}, and its abundant presence has stimulated in the last decades a very active area of research. The Kuramoto model \cite{kuramoto1975,strogatz2000,pikovsky2001,acebron2005} has been used as one of the most versatile tools to understand the different scenarios  where a population of heterogeneous units can develop a common rhythm through mutual interaction, despite the intrinsic tendency of each element to oscillate with its own natural frequency when uncoupled from the network.
In general, the units   in question are located on the vertices of a complex undirected network \cite{newman2010}, which is described by its adjacency matrix $A_{ij}$, with binary entries $A_{ij} = 1$ if there is an edge between vertices $i$ and $j$ and $A_{ij} = 0$ otherwise. The Kuramoto model is then defined  by the nonlinear system of differential equations \cite{arenas2008}
\begin{equation}
\frac{d \theta_{i}}{dt} = \omega_i + \lambda \sum_{j=1}^{N} A_{ij} \sin(\theta_j - \theta_i),
\label{kuramoto_model}
\end{equation}  
where the phase $\theta_{i}(t)$ corresponds to the state of the $i$th unit, 
 which would oscillate with its natural frequency $\omega_i$ if uncoupled from the network. The interaction between connected units is governed by the sine of their phase difference, and the strength of the coupling is determined by the parameter $\lambda$. 
Typically, in order to mimic the inherent differences between the elements of real systems, the natural frequencies $\omega_i$ are 
assumed to be random variables with a  distribution $g(\omega)$, which  we will consider here to be symmetric and unimodal.  

A convenient way to describe the global state of the Kuramoto oscillators (\ref{kuramoto_model}) is to use the  order parameter defined as
\begin{equation}
re^{i\psi} = \frac{1}{N}\sum_{j=1}^{N} e^{i\theta_j},
\label{kuramoto_order_parameter}
\end{equation}
where $r$ is assumed to be real, 
which corresponds to the centroid of the phases if they are considered as a swarm of points moving around the unit circle.  For incoherent motion, the phases are scattered on the circle homogeneously and $r \approx N^{-1/2}$, while for a synchronized state they should move in a single {lump} and, consequently, $r \approx 1$. The Kuramoto model is known to exhibit a second order phase transition from the  incoherent to the synchronized regime at a critical value $\lambda_{c}$ of the coupling strength. For $\lambda \ge \lambda_{c}$, the order parameter $r$ is an increasing function of $\lambda$.
A very natural question to set forth here is: given $N$ vertices with natural frequencies $\left\{ \omega_i \right\}_{i=1}^{N}$ and $m$ edges, what is the best way to connect the vertices (forming a connected network) in order to optimize synchronization (maximize $r$)?  In the last years, it has become clear \cite{brede2008a,buzna2009,kelly2011,skardal2014} that optimal (in the synchronization sense) networks of Kuramoto oscillators typically have a strong negative correlation between the natural frequencies of adjacent oscillators, and a positive correlation between the frequency magnitude $|\omega_i|$ and the degree $k_{i} = \sum_{j=1}^{N} A_{ij}$ of the vertex where the $i$ oscillator lies. This situation   favors some heterogeneity in the degree distribution: vertices with large positive (negative) natural frequencies tend to have higher degrees and to be surrounded by vertices with negative (positive) natural frequencies. These properties are also observed \cite{pinto2014} in the optimized networks for the Kuramoto model with inertia \cite{filatrella2008}, although in this case the optimization has more severe consequences, as it also changes the nature of the phase transition: optimized networks typically possess a first order transition  instead of the usual second order one.

We notice also that many recent works have been  devoted   to the optimization of the synchronization for the case of identical oscillators with diffusive coupling in a complex network. For such a case, the configuration where all the oscillators behave identically is a solution of the equations of motion, albeit not necessarily a stable one. In order to guarantee the stability of this synchronized state \cite{pecora1988,barahona2002}, one finds that the ratio $R = \varpi_N /\varpi_2$ between the largest eigenvalue $\varpi_N$ and the smallest non-zero eigenvalue $\varpi_2$ of the Laplacian matrix $L_{ij} = \delta_{ij}k_i - A_{ij}$ must be smaller than a certain value, which, on the other hand, is uniquely
 determined by the internal dynamics of the vertices (independent of the network  topology, see also \cite{m1,m2,m3}). Moreover,  the smaller the ratio $R$, the better is the stability of the synchronized state. There has been a great amount of work in studying the properties of networks obtained from the minimization of the ratio $R$ trough some kind of rewiring \cite{donetti2005},
including in the case of   multiplex networks \cite{dwivedi2015}. We note, however, that these approaches are not 
suitable    when the natural frequencies are random variables as in the cases considered here.   

In this work, we revisit   the issue of optimization of synchronization  in Kuramoto networks, but  employing  the
dimensional reduction approach recently proposed by Gottwald \cite{gottwald2015}, which explores some tools of the theory of solitary waves
 by introducing    a collective coordinate for the time evolution of the phase locked oscillators, 
in the spirit of the Ott-Antonsen ansatz \cite{ott2008,ott2009}.  Thanks to  this dimensional reduction, we are able to derive analytically a simple condition for optimizing the network topology to favor synchronization in the Kuramoto model (\ref{kuramoto_model}).
We will show that the optimal synchronization of (\ref{kuramoto_model}) for $\lambda>\lambda_c$ demands the minimization of the quadratic
function
\begin{equation}
f'(0) = \sum_{i=1}^{N} \sum_{j=1}^{N} A_{ij}  \omega_i \omega_j - \sum_{i=1}^{N} k_{i} \omega_{i}^{2} = -\omega^TL\omega,
\label{opt_condition_kuramoto}
\end{equation}
where $\omega$ is the vector formed by the natural frequencies of the oscillators, and $L$ the Laplacian matrix. 
Many previously known results can be readily obtained from the minimization of (\ref{opt_condition_kuramoto}) or, in other
words, from the maximization of $\omega^TL\omega$. Moreover, our condition can be adapted into a hill climb algorithm to produce optimal networks in a very efficient way, since it is not necessary the integration of any differential equation or the computation of  any matrix eigenvector
and  our condition involves only   only  plain matrix multiplications.
We will also apply  Gottwald's dimensional reduction approach  \cite{gottwald2015} to study the optimal networks for the Kuramoto model with attractive and repulsive interactions \cite{hong2012}
\begin{equation}
\frac{d \theta_{i}}{dt} = \omega_i + \sum_{j=1}^{N} A_{ij} \lambda_{j} \sin(\theta_j - \theta_i),
\label{kuramoto_model_with_contrarians}
\end{equation}  
where the coupling strengths $\lambda_{j}$ can now be either positive or negative, encoding not only the strength, but also the
 sign of the influence of
oscillator $j$ on its neighbors. Positive values of $\lambda_{j}$ promote   in-phase relationships between neighbors, whereas negative values do   anti-phase ones.
For this kind of model, we also show that
optimal synchronization demands the maximization of a quadratic function involving the
Laplacian matrix associated to the weighted adjacency matrix ${\cal A} = A\Lambda$ , with $\Lambda_{ij} = \delta_{ij}\lambda_j$. 
Our results in this
 case  
 are also  compatible
with those ones   recently discussed in \cite{louzada2012,li2013,zhang2013}.

In the next section, we will derive our results for the Kuramoto model (\ref{kuramoto_model}), introduce our main algorithm, and discuss an explicit example. Section III is devoted to the extension of our approach to  Kuramoto models with attractive and repulsive interactions (\ref{kuramoto_model_with_contrarians}), while the last section is left to some final remarks.

\section{Optimal synchronization in the  Kuramoto model}
\begin{figure}[ht]
\centering
\includegraphics[scale=0.2]{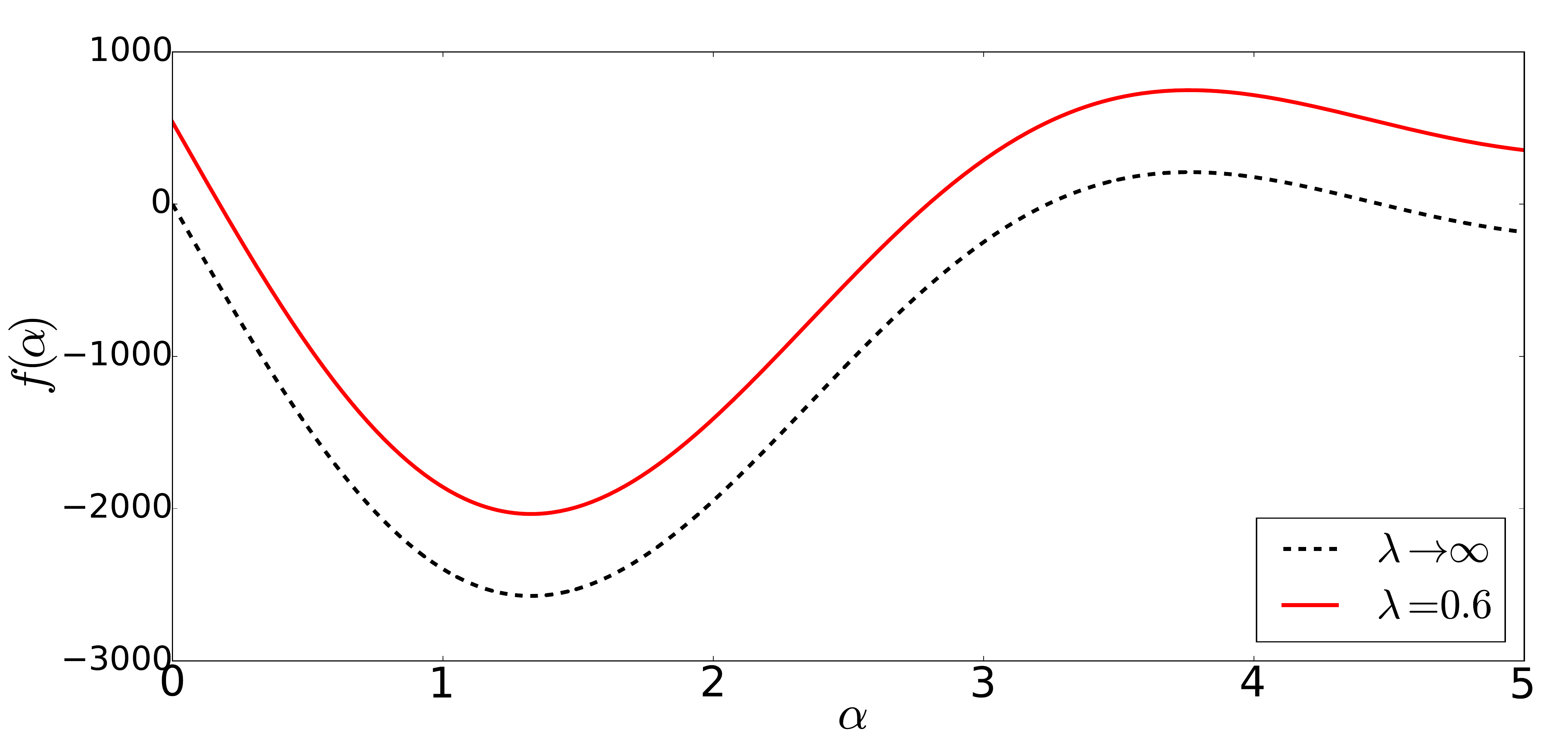}
\caption{(Color online). The graphics show the function $f(\alpha)$ (\ref{gottwald_equation2}) for an Erdos-Renyi network with $N = 1000$ vertices, mean degree $\langle k \rangle = 10$ and natural frequencies drawn from the uniform distribution in the interval $(-1,1)$, for two different values of the coupling strength $\lambda$.}
\label{gottwald_equation_figure}
\end{figure}

Since the   Kuramoto model (\ref{kuramoto_model}) has rotational invariance, we can change, without loss of generality, to a new   reference frame 
$\theta_i \to \theta_i + \Omega t$
such that the distribution $g(\omega)$ is centered at $\omega = 0$ (which, in our case, implies $\langle \omega\rangle =0$). 
Several well known results suggest  that for the Kuramoto model (\ref{kuramoto_model}), the time evolution of the phase locked oscillators may be well approximated by using the collective parametrization \cite{gottwald2015}
\begin{equation}
\theta_{i}(t) = \alpha(t) \omega_{i}.
\label{kuramoto_parametrisation}
\end{equation}
By using such  ansatz  in the equations of motion (\ref{kuramoto_model}) in the
reference frame for which $\langle \omega\rangle =0$,  and demanding that a certain error function is minimal, Gottwald end up with a one-dimensional differential equation for $\alpha$ \cite{gottwald2015}
\begin{equation}
\dot{\alpha} = \frac{\lambda}{\sigma^2}f(\alpha)  ,
\label{gottwald_equation}
\end{equation}
with 
$
\sigma^2 = \sum_{i=1}^{N} \omega_{i}^2, 
$
and
\begin{equation}
f(\alpha) =   \frac{\sigma^2}{\lambda} + \sum_{i=1}^{N} \omega_{i} \sum_{j=1}^{N} A_{ij} \sin(\alpha(\omega_j - \omega_i)).
\label{gottwald_equation2}
\end{equation}
With the ansatz (\ref{kuramoto_parametrisation}), synchronized solutions correspond to stable fixed points $\alpha^{*}$ of (\ref{gottwald_equation}), as the difference of phases $\alpha(t) \left( \omega_j - \omega_i \right)$ are constants  if $\dot{\alpha} = 0$. Hence, 
the fixed points $\alpha^{*}$ must satisfy  the condition $f(\alpha^{*}) = 0$, and its stability  is determined
by the sign of $f'(\alpha^*)$. 
Figure (\ref{gottwald_equation_figure}) 
shows $f(\alpha)$ for two different values of $\lambda$,  $\lambda \rightarrow \infty$ and $\lambda = 0.6$, for an Erdos-Renyi network with $N = 1000$ vertices, mean degree $\langle k \rangle = 10$ and natural frequencies drawn from the uniform distribution in the interval $(-1,1)$. For the first case, we clearly see that there is a stable fixed point $\alpha^{*} = 0$, as one would expect. The curve for $\lambda = 0.6$, however, is the same one for $\lambda \rightarrow \infty$, but shifted upwards, displacing the stable fixed points towards  right.

This scenario allows us to devise a method to optimize the synchronization for finite values of $\lambda$. We want
to have $r$   as large as possible for a given finet value of $\lambda>\lambda_c$. 
 In order to increase the order parameter $r$, we must assure that the phase's difference in the synchronized state are as small as possible, {\em i.e}, $\alpha^{*}$ must be minimal, so that every oscillator is close to each other accordingly to (\ref{kuramoto_parametrisation}). If we linearize the function $f(\alpha)$ around $\alpha = 0$ and demand that it crosses the abscissa as close as possible to the origin, we obtain the condition that   
$f'(0)$ must be minimal, which, according to (\ref{opt_condition_kuramoto}) implies that
$\omega^TL\omega$ must be maximal. 
 Notice that, interestingly, this conditions does not depend on the value of $\lambda$.  It is clear also  that
optimal synchronization for the Kuramoto model demands   a negative correlation of the natural frequencies of adjacent oscillators and a positive correlation between the degrees $k_{i}$ and the values of $|\omega_{i}|$, 
which  are precisely the  results obtained  previously by different and   more intricate numerical approaches \cite{brede2008a,buzna2009,kelly2011,skardal2014}.

A {hill climb} rewiring algorithm can be easily set up to find the maximum  of $\omega^TL\omega$ and, hence, to produce optimal
synchronization networks. The strategy is roughly the following. A randomly selected edge connecting two vertices is removed if it does not disconnect the network, and two randomly chosen disconnected vertices are then connected. After this step, the new value of $\omega^TL\omega$ is evaluated. If the rewiring results in a higher value, one keeps the modification or, otherwise, one discharges the rewiring and returns the network to its previous configuration. This procedure is repeated until $\omega^TL\omega$ attains a minimum value, what of course will guarantee
a minimal value for $f'(0)$. In practice, our algorithm limits the maximum number of iterations (up to $2\times10^4$ times in our simulations). These edge swaps preserve the average degree of the initial network since the number of edges is kept the same, but not the degree distribution.   
It is clear that this kind of hill climb algorithm works by performing a local search of the optimal state by incrementally changing the structure of the network, and it is indeed the simplest algorithm to perform the   maximization of
$\omega^TL\omega$. Compared with  more complex algorithms such as simulated annealing \cite{kirkpatrick1983}, for instance,
our hill climb algorithm has proved to be much faster and extremely reliable. However,  there are some potential issues, as for instance,   the algorithm might get stuck in a local 
minimal 
 state far from the global minimum. Nevertheless, our numerical simulations show that this is very rare and we can indeed get networks with pronounced enhancement of the synchronization capacity from this simple algorithm, which is indeed the most used one in the literature of optimization of synchronization in complex networks (see, for instance, \cite{brede2008a,buzna2009,kelly2011}).

\begin{figure}[ht]
\centering
\includegraphics[scale=0.2]{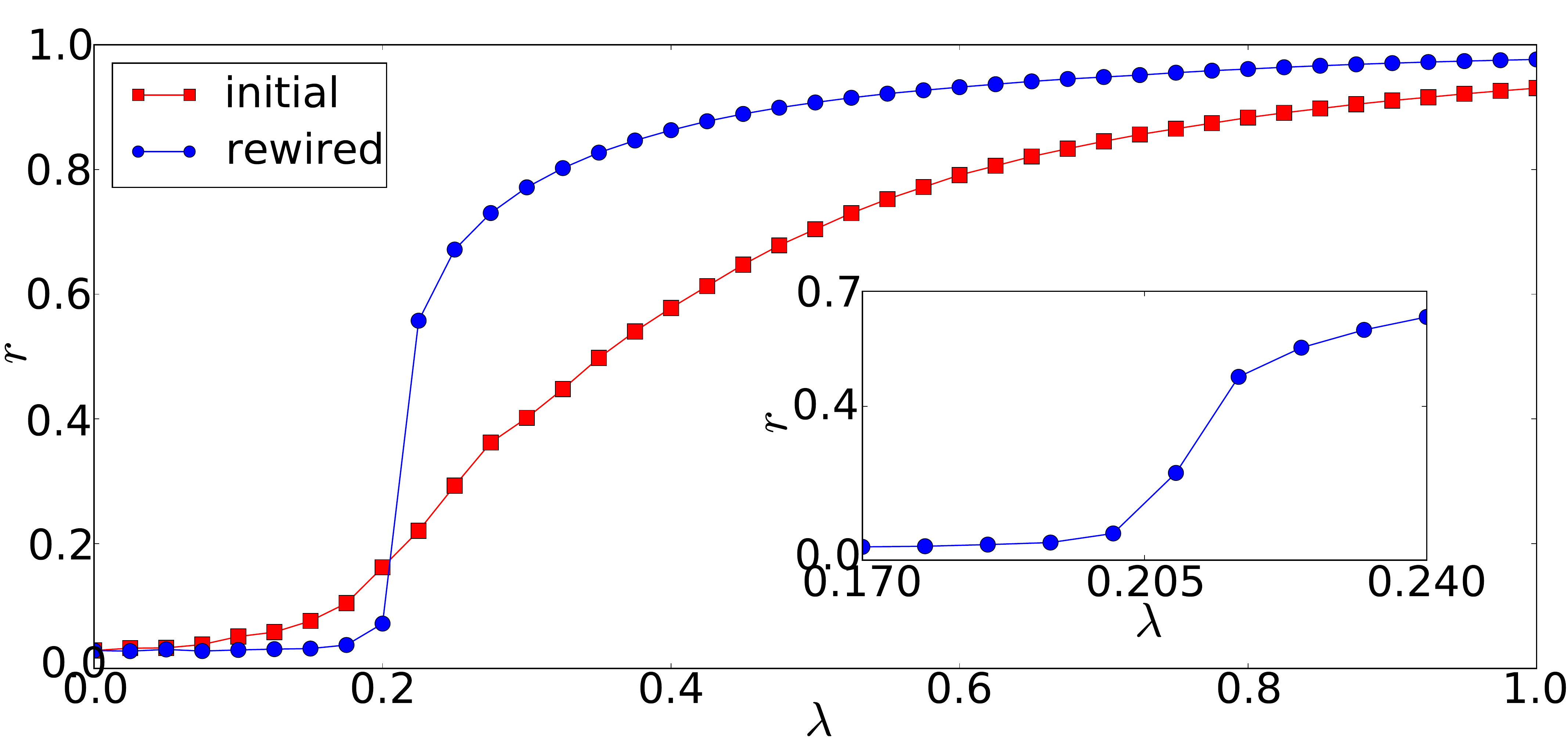}
\caption{(Color online). Synchronization diagrams for the initial and optimized networks for the Kuramoto model (\ref{kuramoto_model}). The diagram for the initial network, a Barabasi-Albert network with $N = 10^{3}$ vertices and $\langle k \rangle = 6$, is the curve with the red squares. The natural frequencies were drawn from the unit normal distribution. The optimized network, shown as the curve with blue circles, was obtained from the maximization  of $\omega^TL\omega$ by using the \emph{hill climb} algorithm described in the text. The inset shows the region around the phase transition for the rewired network, confirming that is of the second order type.}
\label{ba_phase_diagrams}
\end{figure}

\begin{figure}[ht]
\centering
\includegraphics[scale=0.2]{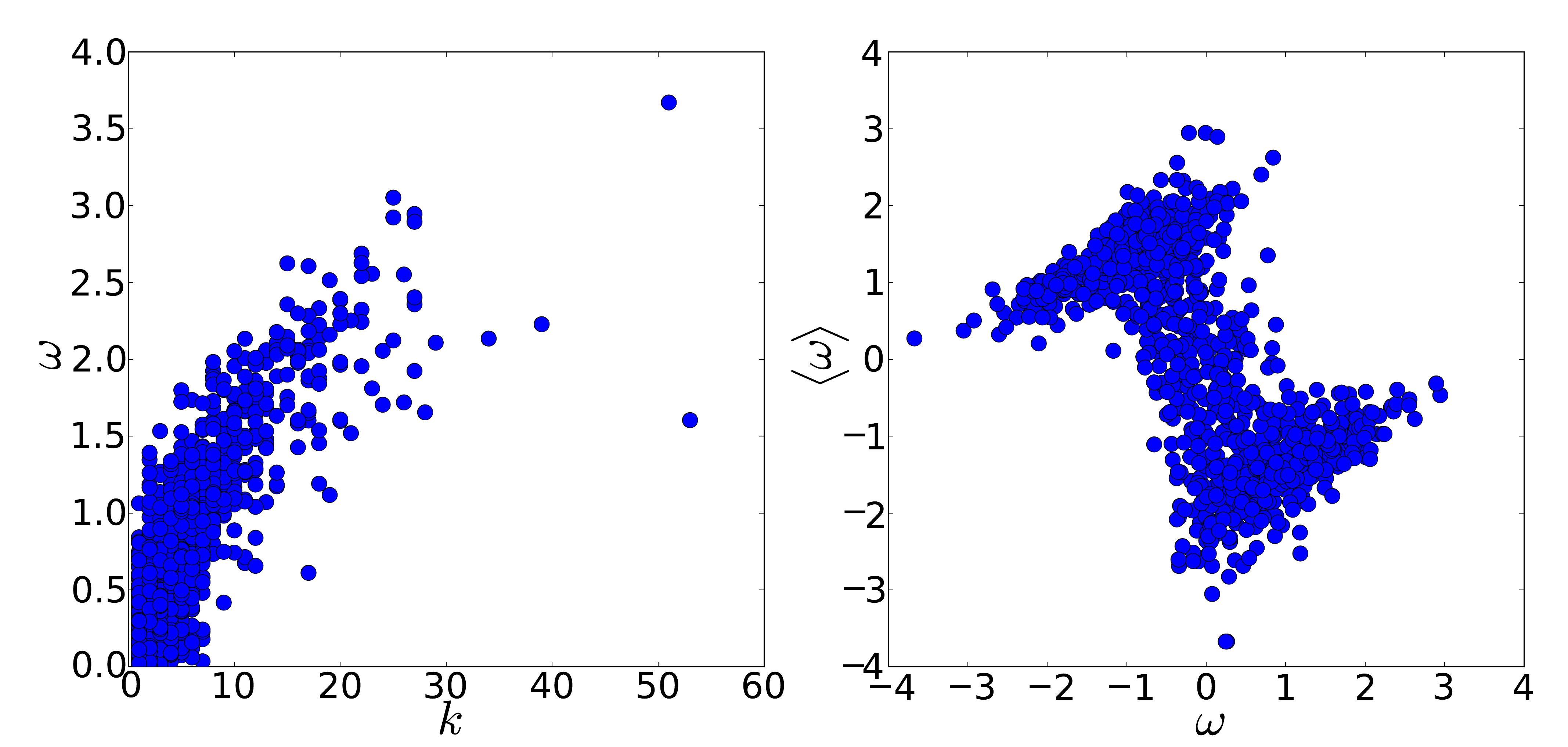}
\caption{(Color online). The graphics shows in the left panel the positive correlation between natural frequencies $|\omega|$ of the oscillators and degrees $k$ of the vertices where they lie. The negative correlation between natural frequencies of adjacent vertices, measured here by the average natural frequency $\langle \omega_i\rangle_{\cal N}=\sum_{j=1}^{N} A_{ij}\omega_{j} /k_{i}$ of the neighbours ${\cal N}_i$ of a vertex $i$ as a function of its natural frequency $\omega_i$, is 
depicted in the right panel. }
\label{opt_correlations}
\end{figure}

As an explicit example, we apply this algorithm  for a network built with the Barabasi-Albert method \cite{barabasi1999} (which creates scale free networks with degree distribution $p(k) \propto k^{-3}$) with $N = 10^{3}$ vertices and mean degree $\langle k \rangle = 6$. The natural frequencies were drawn from the unit normal distribution. The synchronization diagram for this network is shown as red squares in Fig. \ref{ba_phase_diagrams},
where the usual smooth monotonically increasing behavior for $r$ over all the range of coupling strength $\lambda$ is observed.
The synchronization diagram of the optimized network that results from the algorithm is shown in Figure \ref{ba_phase_diagrams} as blue circles. It   is evident the considerable  enhancement  of the network synchronization capacity over a large range of coupling strengths. We stress  that the phase transition is now   much more abrupt, but still of second order, since the zoom around the critical region shown in the inset has a continuous behavior, and no hysteresis loop seems to be present. 
Figure \ref{opt_correlations} 
shows   the positive correlation between natural frequencies $|\omega|$ and degrees $k$   for the optimized networks (left panel), as well as the negative correlation between the natural frequencies of adjacent oscillators (right panel). We can give yet another measure of this strong negative correlation noticing  that the fraction of links connecting oscillators with positive and negative frequencies jumps from $0.5$ to $0.87$ in this particular run of the optimization procedure.

The condition of maximal $\omega^TL\omega$ allows us also to attack  another related problem considered recently in the literature.
The Laplacian matrix $L$ of an undirected network is symmetric, so that all its eigenvalues are real. Furthermore it can be demonstrated \cite{newman2010} that for a connected network it has only one eigenvalue $\varpi_1 = 0$ and all the other $N - 1$ are positive.  Suppose that the topology of the network is fixed and we want to obtain the set of natural frequencies to be allocated to each vertex such that it optimizes the synchronization properties, subjected to the
constraint  $\sigma^2 =  \sum_{i=1}^{N} \omega_{i}^{2}  = N$ (in fact, any positive constant). By writing $\mathbf{\omega}$ as a linear combination of the eigenvalues $\mathbf{v}_i$ of $L$, $\mathbf{\omega} = \sum_{i=1}^{N} \alpha_{i} \mathbf{v}_i$, where $\sum_{i=1}^{N}\alpha_{i}^{2} = N$ 
to assure the validity of constraint, we have that in order to maximize
$\omega^TL\omega$ one must set $\mathbf{\omega}$ proportional to the eigenvector of largest eigenvalue of $L$ ($\alpha_{1} = ... = \alpha_{N-1} = 0$ and $\alpha_{N} = \sqrt{N}$), a result recently found in \cite{skardal2014}  by using more intricate  methods.

\section{The Kuramoto model with attractive and repulsive interactions}

\begin{figure}[ht]
\centering
\includegraphics[scale=0.2]{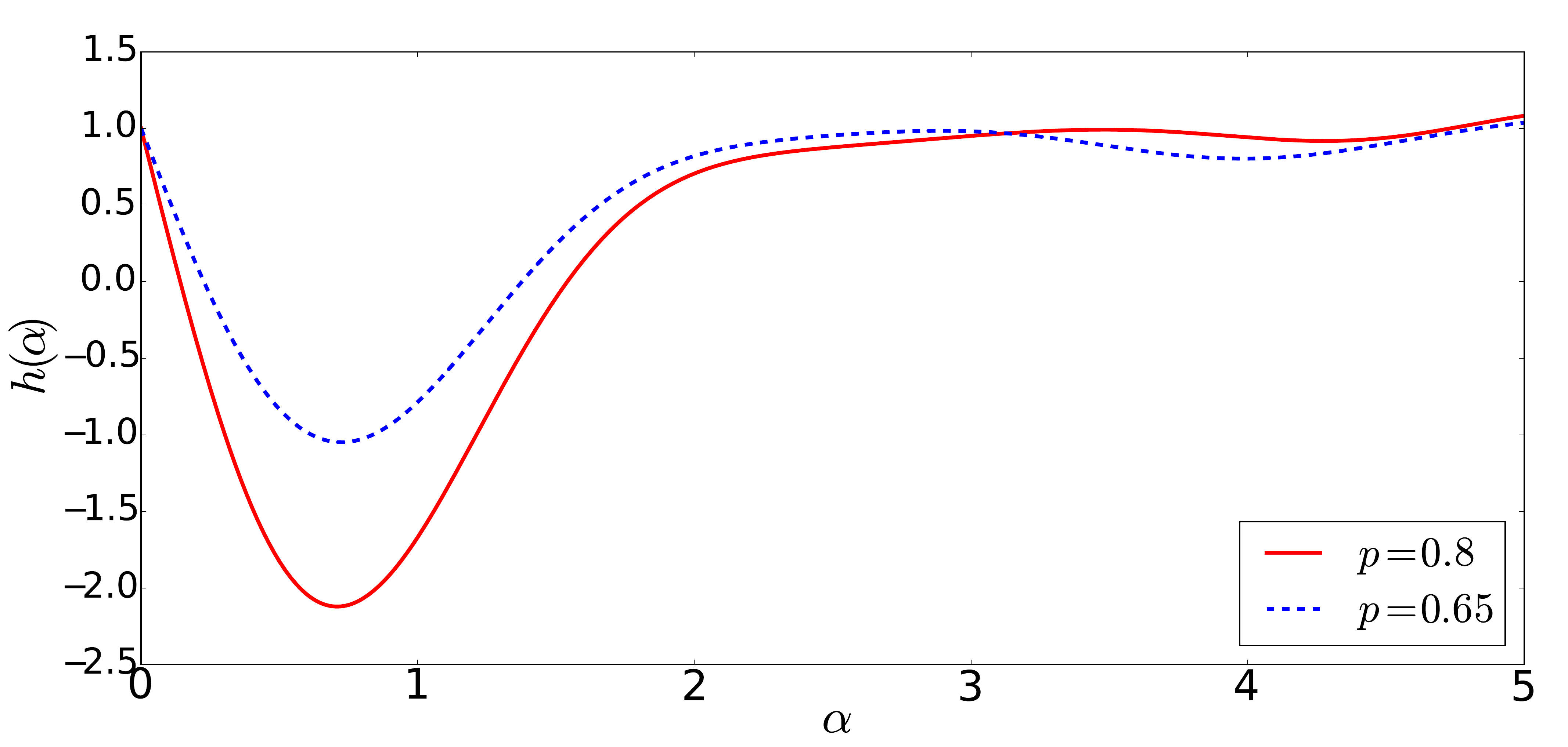}
\caption{(Color online). The graphics show the function $h(\alpha)$ (\ref{h_function}) for an Erdos-Renyi network with $N = 10^3$ vertices and natural frequencies drawn from the unit normal distribution, for two different values of the fraction $p$ of vertices with positive coupling strength. For both curves, $\lambda^{-} = -0.5$.}
\label{gottwald_equation_contrarians_figure}
\end{figure}

For the Kuramoto model with attractive and repulsive interactions (\ref{kuramoto_model_with_contrarians}), we will take advantage of the
rather unexpected result that for both kinds of oscillator, those with positive and negative couplings, the collective parametrisation of the phase locked oscillators can be taken to be the same, see, for instance, Figure 3 of \cite{hong2012}. Hence, we can employ the same ansatz (\ref{kuramoto_parametrisation}) also in this case. 
We will consider the situation with two kinds of units: the attractive and repulsive ones, for which the
coupling strength are, respectively, $\lambda^{+}$ and $\lambda^{-}$.
Let $p$ be the fraction   of attractive oscillators, and, consequently, $1-p$ the fraction of repulsive ones.   As the coupling strengths are properties of each oscillator, it is possible that an oscillator with positive coupling strength is connected to an oscillator with negative strength, causing in this way a kind of frustration effect in the network. 
Without loss of generality, one can rescale the time variable in order to have $\lambda^{+} = 1$.  An issue that arises here is that the frequency of rotation of the locked oscillators is not equal to the average value of the natural frequencies $\langle \omega \rangle$, as it happens for the Kuramoto model (\ref{kuramoto_model}). By changing to a new reference frame
 $\theta_i\to\theta_i+\Omega t$, one has from (\ref{kuramoto_model_with_contrarians})
 \begin{equation}
\frac{d \theta_{i}}{dt} = \omega_i - \Omega + \sum_{j=1}^{N} A_{ij} \lambda_{j} \sin(\theta_j - \theta_i).
\end{equation} 
Multiplying both sides by $\lambda_{i}$ and then summing with respect to the index $i$, the term $\sum_{i=1}^{N} \sum_{j=1}^{N} A_{ij} \lambda_{i} \lambda_{j} \sin(\theta_j - \theta_i)$ cancel out, and we have that a synchronized state must have
\begin{equation}
\Omega = \frac{ \sum_{i=1}^{N} \lambda_{i} \omega_{i} } { \sum_{i=1}^{N}\lambda_{i} }.
\label{freq_phase_locked}
\end{equation}
Repeating here the same steps of Gottwald \cite{gottwald2015}, 
we will have that $\alpha(t)$  must obey, in this case, the ordinary differential equation 
$
\dot{\alpha} = h(\alpha),
$ where
\begin{equation}
h(\alpha) = 1 + \frac{1}{\sigma^2} \sum_{i=1}^{N} \omega_i \sum_{j=1}^{N} A_{ij} \lambda_{j} \sin(\alpha(\omega_j - \omega_i)).
\label{h_function}
\end{equation}
In the same manner as for the Kuramoto model, to obtain the stable synchronized solutions we must seek values of $\alpha^{*}$ such that $h(\alpha^{*}) = 0$ with $h'(\alpha^*)<0$. 
Two examples of the function $h(\alpha)$ are shown in Figure \ref{gottwald_equation_contrarians_figure} 
for an Erdos-Renyi network with $N = 10^3$ vertices, mean degree $\langle k \rangle = 10$ and natural frequencies drawn from a unit normal distribution. The curves correspond to two different values of the fraction $p$ of oscillators with positive coupling strengths (and $\lambda^{-} = -0.5$).
From the overall shape of the function $h(\alpha)$, the condition for the optimization, either to favor or to suppress synchronization, depends on the value of the derivative of $h(\alpha)$ at $\alpha = 0$,
\begin{equation}
\sigma^2 h'(0) =  \sum_{i=1}^{N}  \sum_{j=1}^{N} A_{ij}\lambda_j \omega_{i}\omega_j  - \sum_{i=1}^{N}    \sum_{j=1}^{N} A_{ij} \lambda_{j} \omega_{i}^{2}  = -\omega^T {\cal L}\omega ,
\label{alpha_0_cond}
\end{equation}
where $\cal L$ is the Laplacian matrix associated to the weighted adjacency matrix ${\cal A} = A\Lambda$, 
\begin{equation}
{\cal L}_{ij} = \delta_{ij}\sum_{k=1}^N {\cal A}_{ik} - {\cal A}_{ij},  
\end{equation}
with $\Lambda_{ij}=\delta_{ij}\lambda_j$. Clearly,  when $\lambda_{i} = \lambda$ for all oscillators, we recover the previous 
optimization condition. In contrast with the previous case, the  Laplacian matrix ${\cal L}$  is not symmetrical here,    rendering the 
spectral 
analysis much more complicate in this case. 

However, 
the overall scenario is very similar to that one found for the standard Kuramoto model, and we will proceed in the same manner to optimize the network structure either to favor or to suppress synchronization. Of course, for the first case, we must 
maximize $\omega^T {\cal L}\omega$, whereas for the second case we must minimize it.
It is interesting that   we also have here a rather clear interpretation of the characteristics of the optimized network in terms of correlations between microscopic properties. Inspecting equation (\ref{alpha_0_cond}), we see that if we are optimizing the network to favor synchronization, there must be a  negative correlation between the natural frequency $\omega_{i}$ of the oscillator placed at vertex $i$ and the average value of the product $\omega_{j}  \lambda_{j}$ over its neighbors, as well as a positive correlation between $|\omega_{i}|$ and the average value of coupling strengths of its neighbors. If we are optimizing the network to suppress synchronization, the correlations must be reversed.
The same hill climb rewiring  algorithm can be implemented to minimize or maximize condition  $\omega^T {\cal L}\omega$ and produce networks with the desired properties.

\begin{figure}[ht]
\centering
\includegraphics[scale=0.2]{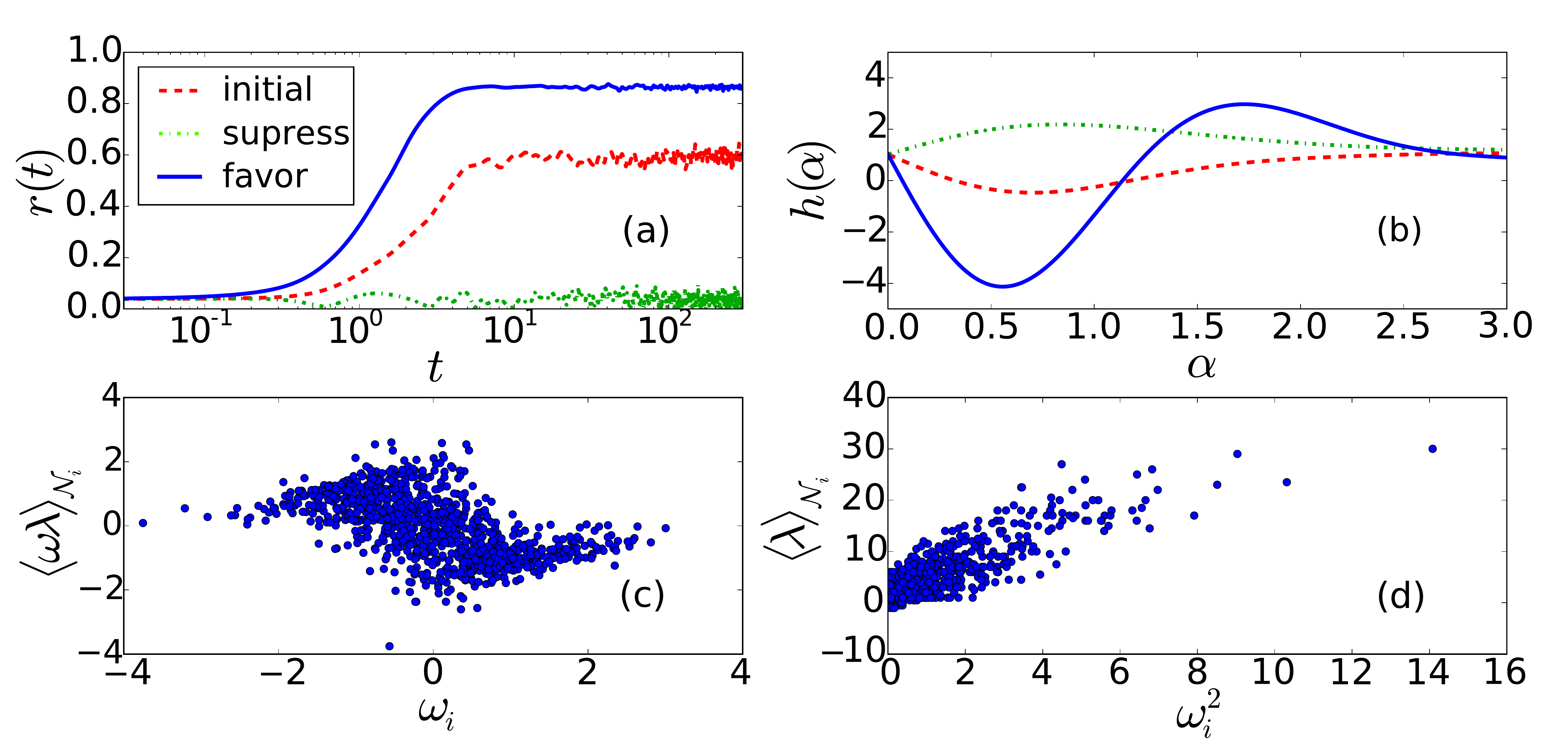}
\caption{(Color online). Panel (a) shows the time evolution of $r(t)$ for the original network (the red dashed line), for the optimized network in order to favor synchronization (the blue continuous line) and for the optimized network in order to suppress synchronization (the dash-dotted green line). Panel (b) shows the behavior of function $h(\alpha)$ for the three cases depicted in panel (a). Panels (c) and (d) show, respectively, the negative correlation between the natural frequency $\omega_{i}$ of the oscillator at vertex $i$ and the mean value of the product $\langle \omega \lambda \rangle$ over its set of neighbors $\mathcal{N}_{i}$ and the positive correlation between $\omega_{i}^{2}$ and the mean value of the coupling $\langle \lambda \rangle$ of $\mathcal{N}_{i}$ for the optimized network to favor synchronization shown in panel (a).}
\label{opt_contrarians_figure}
\end{figure}

\begin{figure}[ht]
\centering
\includegraphics[scale=0.2]{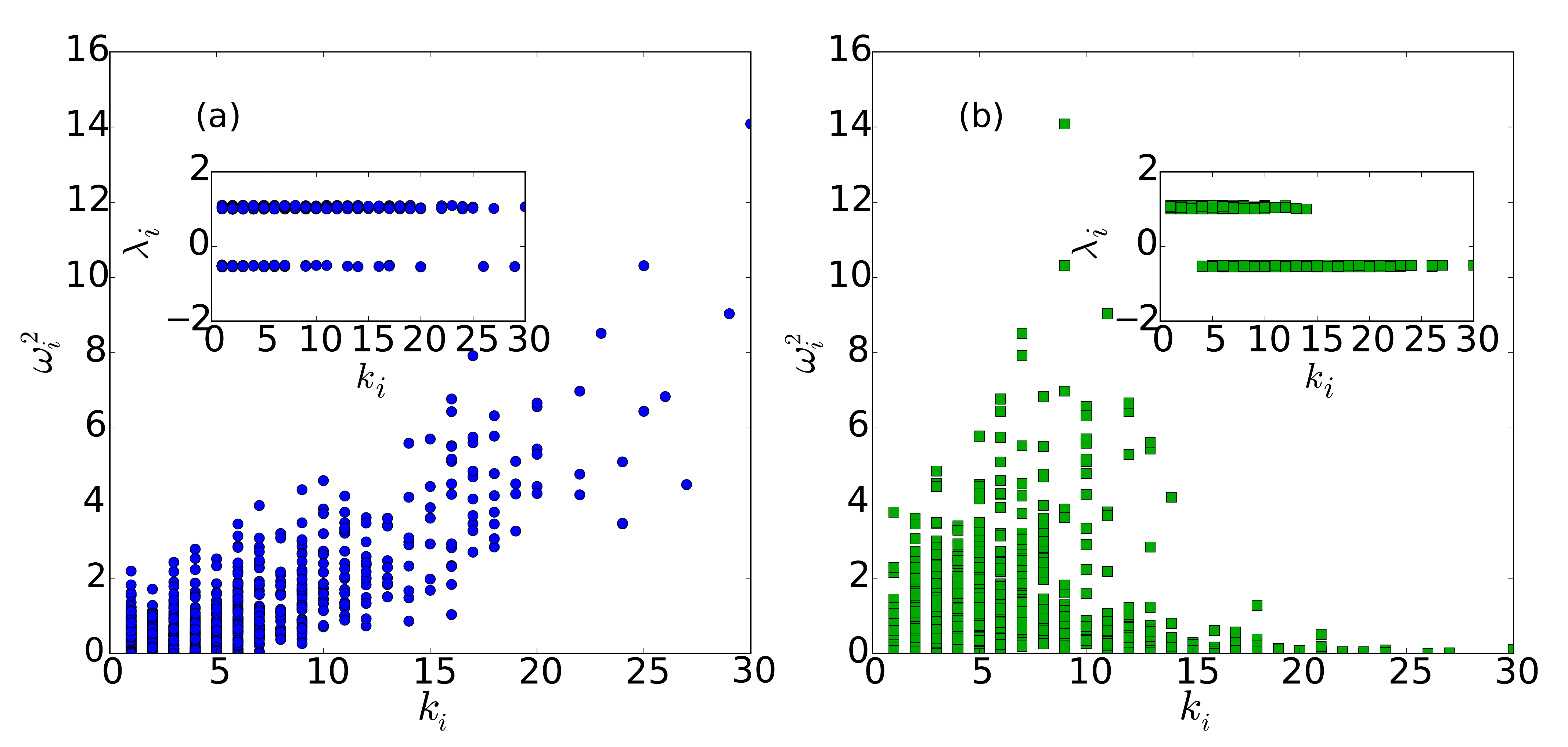}
\caption{(Color online). The graphics show the positive correlation between the magnitude of the natural frequencies and degrees for the optimized network to favor synchronization, panel (a), whereas there is no correlation for the network optimized to suppress synchronization, panel (b). Both results correspond to the networks depicted in figure \ref{opt_contrarians_figure}. The insets show, for both cases,
the distribution of the couplings as function of the degrees.    }
\label{contrarians_versus_degree}
\end{figure}

We apply our algorithm  to build the optimal network topologies, both to favor and to suppress synchronization, using as seed an Erdos-Renyi network with $10^{3}$ vertices and mean degree $\langle \omega \rangle = 5$. The natural frequencies were drawn from the unit normal distribution and a fraction $p = 0.8$ of the vertices have positive strengths
$\lambda^+=1$, while $0.2$ of the vertices have  $\lambda^{-} = -0.5$. The natural frequencies and coupling strengths for each vertex were randomly assigned and were kept fixed during the optimization procedure, the algorithm only performs the rewiring of  the network connections. 
The results are shown in Figure \ref{opt_contrarians_figure}. 
Panel (a) shows the time evolution of the order parameter $r(t)$ for the original network (the red dashed line), for the optimized network in order to favor synchronization (the blue continuous line) and for the optimized network in order to suppress synchronization (the dash-dotted green line). The overall behavior of the three cases are independent of the initial conditions.
The method works very well for both cases, the synchronization properties are clearly enhanced for the optimal synchronization network, while
the   network optimized to suppress synchronization shows only a noise-like behavior of the order parameter $r$.
The panel (b) shows the functions $h(\alpha)$ for the three cases depicted in panel (a). We can see that for the optimized network to suppress synchronization, the derivative $h'(0)$ reversed its sign, destroying the stable synchronized solution. Panels (c) and (d) show, for the network optimized to favor synchronization, the presence of the correlations discussed earlier, respectively, the negative correlation between the natural frequency $\omega_{i}$ of the oscillator at vertex $i$ and the mean value
\begin{equation}
\langle \omega \lambda \rangle_{\cal N} = \frac{\sum_{j=1}^N A_{ij}\omega_j\lambda_j}{\sum_{j=1}^N A_{ij} \lambda_j }
\end{equation}
 over its set of neighbors $\mathcal{N}_{i}$, and the positive correlation between $|\omega_{i}|$ and the mean value of the coupling $\langle \lambda \rangle$ of its neighbors $\mathcal{N}_{i}$. 
We have found also a positive correlation between the magnitude of the natural frequencies and degrees for the optimized network to favor synchronization, see Figure \ref{contrarians_versus_degree}, 
whereas no correlation (either positive or negative) is seem for the optimized network to suppress synchronization. 
Another interesting result is that for the optimal synchronization case, the positive and negative coupling strengths are equally distributed in a large range
of degrees, while for the case of optimized network to suppress synchronization, most of the oscillators with negative coupling are located at large degree vertices (see the insets of Figure \ref{contrarians_versus_degree}).
Notice that the optimized network to favor synchronization does not show $r \approx 1$, it instead  saturates at the lower value $r \approx 0.9$
due to the  already commented frustrating effects associated with the combination of attractive and negative units.

\section{Conclusions}

We have shown  that the  dimensional reduction approach recently proposed by Gottwald  \cite{gottwald2015}  for the Kuramoto model can be adapted
 to obtain   a simple analytical condition  to optimize the network topology to favor or to suppress synchronization. 
Our condition allowed us to rederive analytically some recent results \cite{brede2008a,buzna2009,kelly2011,skardal2014}
obtained numerically. Our approach was also 
extended  to the Kuramoto model with attractive and repulsive interactions, leading also to  
 simple analytical condition  to optimize the networks, complementing in this way  some recent works
  \cite{louzada2012,li2013,zhang2013} which considered the optimization of synchronization  in these models numerically. 
  In both cases, the optimal synchronization condition corresponds to the maximization of the quadratic funcion $\omega^TL\omega$, where
 where $\omega$ stands for the vector of 
the
 natural frequencies of the oscillators, and $L$ for the pertinent network Laplacian matrix.
The optimization condition involves only   
the microscopic properties of the network,
enlightening  in this way many correlations observed numerically for optimal synchronization networks. We could 
introduce a hill climb rewiring algorithm to produce optimized networks in a very efficient way, since it is not necessary the integration of any differential equation or the computation of   any matrix eigenvector, only plain matrix multiplications  are used in each step.
As common for this kind of algorithm, one cannot assure in principle that the procedure will stop effectively in a global minimum. However, since the algorithm is computationally simple, on can run the procedure for several initial networks in order to seek for a global minimum. In all tests we have performed, our algorithm returned, with little computational effort, networks with greatly enhanced synchronization properties.
An important open problem  is the optimization of other kinds of oscillators as, for example, the Winfree model \cite{winfree1967}.  It would be interesting to investigate if the Gottwald approach  \cite{gottwald2015}, or some variant of it, could be applied to these cases, 
and if an   analysis analogous to the presented here could be indeed used to optimize the network topology to favor synchronization or other kinds of collective behaviors that might emerge in these more complicated models.

\section*{Acknowledgements}
AS thanks FAPESP (grant 2013/09357-9) and CNPq for the financial support. RSP thanks CAPES 
for the financial support and G. Gotwald for helpful conversations.  We also thank the referees for the suggestions and
 criticisms. 
Our numerical computations were done by using the SciPy package for python
\cite{SciPy} and the NetworkX
package \cite{NetworkX}.

\end{document}